\documentclass{article}
\usepackage[accepted]{mlsys2020}
\usepackage{microtype}
\usepackage{graphicx}
\usepackage{subcaption}
\usepackage{tablefootnote}
\usepackage[flushleft]{threeparttable}
\usepackage{booktabs} 
\usepackage{bm}
\usepackage{amssymb}
\usepackage{amsthm}
\usepackage{amsmath,amssymb,amsfonts}
\usepackage{multicol}
\usepackage{multirow}

\usepackage{hyperref}

\mlsystitlerunning{Towards Scalable Distributed Training of Deep Learning on Public Cloud Clusters}

\begin{document}

\twocolumn[
\mlsystitle{Towards Scalable Distributed Training of Deep Learning on Public Cloud Clusters}




\mlsyssetsymbol{equal}{*}

\begin{mlsysauthorlist}
\mlsysauthor{Shaohuai Shi}{equal,hkust}
\mlsysauthor{Xianhao Zhou}{equal,tencent}
\mlsysauthor{Shutao Song}{equal,tencent}
\mlsysauthor{Xingyao Wang}{michigan}
\mlsysauthor{Zilin Zhu}{tencent}
\mlsysauthor{Xue Huang}{tencent}
\mlsysauthor{Xinan Jiang}{tencent}
\mlsysauthor{Feihu Zhou}{tencent}
\mlsysauthor{Zhenyu Guo}{tencent}
\mlsysauthor{Liqiang Xie}{tencent}
\mlsysauthor{Rui Lan}{tencent}
\mlsysauthor{Xianbin Ouyang}{tencent}
\mlsysauthor{Yan Zhang}{tencent}
\mlsysauthor{Jieqian Wei}{tencent}
\mlsysauthor{Jing Gong}{tencent}
\mlsysauthor{Weiliang Lin}{tencent}
\mlsysauthor{Ping Gao}{tencent}
\mlsysauthor{Peng Meng}{tencent}
\mlsysauthor{Xiaomin Xu}{tencent}
\mlsysauthor{Chenyang Guo}{tencent}
\mlsysauthor{Bo Yang}{tencent}
\mlsysauthor{Zhibo Chen}{tencent}
\mlsysauthor{Yongjian Wu}{tencent}
\mlsysauthor{Xiaowen Chu}{hkbu}
\end{mlsysauthorlist}

\mlsysaffiliation{hkust}{The Hong Kong University of Science and Technology, Hong Kong, China.}
\mlsysaffiliation{tencent}{Tencent Ltd., Shenzhen, China.}
\mlsysaffiliation{michigan}{University of Michigan, Ann Arbor, Michigan, USA, work done while at Tencent.} 
\mlsysaffiliation{hkbu}{Hong Kong Baptist University, Hong Kong, China}

\mlsyscorrespondingauthor{Shaohuai Shi}{shaohuais@cse.ust.hk}
\mlsyscorrespondingauthor{Xianhao Zhou}{jathonzhou@tencent.com}
\mlsyscorrespondingauthor{Shutao Song}{sampsonsong@tencent.com}

\mlsyskeywords{Machine Learning, MLSys}

\vskip 0.3in

\begin{abstract}
Distributed training techniques have been widely deployed in large-scale deep neural networks (DNNs) training on dense-GPU clusters. However, on public cloud clusters, due to the moderate inter-connection bandwidth between instances, traditional state-of-the-art distributed training systems cannot scale well in training large-scale models. In this paper, we propose a new computing and communication efficient top-k sparsification communication library for distributed training. To further improve the system scalability, we optimize I/O by proposing a simple yet efficient multi-level data caching mechanism and optimize the update operation by introducing a novel parallel tensor operator. Experimental results on a 16-node Tencent Cloud cluster (each node with 8 Nvidia Tesla V100 GPUs) show that our system achieves 25\%-40\% faster than existing state-of-the-art systems on CNNs and Transformer. We finally break the record on DAWNBench on training ResNet-50 to 93\% top-5 accuracy on ImageNet.
\end{abstract}
]



\printAffiliationsAndNotice{\mlsysEqualContribution} 

\section{Introduction}
\label{sec:intro}
Due to the increase of deep learning (DL) models and data sets, training models of deep neural networks (DNNs) using stochastic gradient descent (SGD) algorithms, which requires to iteratively update the model parameters to converge, is a compute-intensive process and would be very time-consuming. For example, training a BERT model on a single TPU takes more than 1.5 months~\cite{devlin2019bert}. Distributed training techniques are the common practice to accelerate the training by exploiting multiple processors in a distributed system collaborating on training the model~\cite{dean2012large,goyal2017accurate,you2018imagenet}. 

In the large-scale training with distributed clusters, synchronous SGD with data parallelism is the main training algorithm that has been widely used in industry (e.g., MLPerf\footnote{\url{https://mlperf.org/}}) and academia~\cite{goyal2017accurate,jia2018highly,mikami2018massively,you2020large}. For a $P$-worker distributed system, the $P$ workers collaborate on training the model $\bm{w}_t$ to minimize the objective function $f:\mathbb{R}^d\to \mathbb{R}$ with the following update formula
\begin{equation}\label{equ:ssgd}
    \bm{w}_{t+1}=\bm{w}_t-\eta_t \sum_{p=1}^P \bm{g}_t^p,
\end{equation}
where $\bm{g}_t^p=\nabla f(\bm{w}_t, X_t^p)$ is the stochastic gradient at worker $p$ with sampled data $X_t^p$, and $\eta_t$ is the learning rate. It generally takes tens to hundreds of epochs to converge to a good solution, and one epoch indicates traversing the whole data samples of the data set once. It is of importance to fully exploit the overall computing power to reduce the iteration time to accelerate the training process. According to Eq.~\eqref{equ:ssgd}, each iteration requires the gradient aggregation of distributed workers, which introduces the data communication between GPUs. The gradient aggregation can be implemented with parameter servers~\cite{li2014scaling} and an All-Reduce operation~\cite{baidu2017ring,awan2017s}, among which the All-Reduce operation is more widely used in large-scale training~\cite{goyal2017accurate,you2018imagenet,lin2018deep}. Yet, the gradient aggregation generally introduces high communication overheads compared to the GPU computing time.

On one hand, there exists much work~\cite{jouppi2017datacenter,jia2018highly,wang2018bml,mikami2018massively,ueno2019exhaustive,cho2019blueconnect,wang2020blink,luo2020plink,chu2020nv} in providing performance optimizations for the All-Reduce collective for different environments, but the existing state-of-the-art training systems still scale badly on public cloud clusters that are with fast intraconnects (e.g., NVLink) and slow interconnects (e.g., Ethernet). On the other hand, recently there are many studies on extensive gradient compression techniques that can significantly reduce the communication traffic with very slight loss of accuracy~\cite{alistarh2017qsgd,lin2018deep,shi2019convergence,karimireddy2019error,renggli2019sparcml}. The top-k sparsification algorithm~\cite{lin2018deep} with good convergence guarantees~\cite{stich2018sparsified,alistarh2018convergence} is one of the aggressive algorithms that can only send a small proportion of messages to others with little impact on the convergence.

\begin{table}[!t]
	\centering
    \begin{threeparttable}
	\caption{8 V100 GPUs computing instances on clouds}
	\label{table:public-cloud-instances}
	\addtolength{\tabcolsep}{-2.5pt}
	\begin{tabular}{|c|c|c|c|c|}
	\hline
	 \multirow{2}{*}{Cloud} & \multirow{2}{*}{Instance} & Memory & Storage & Network\\
	  & & (GiB) & Type & (Gbps) \\\hline\hline
	 AWS & p3.16xlarge$^a$ & 488& EBS & 25 \\\hline
	 Aliyun & c10g1.20xlarge$^b$ & 336 &  OSS & 32  \\\hline
	 Tencent & 18XLARGE320$^c$ & 320 & CFS & 25 \\\hline
	\end{tabular}
        \begin{tablenotes}
    	\item \scriptsize{$^a$\url{https://amzn.to/33GQ32w};}
    	\scriptsize{$^b$\url{https://bit.ly/3iRQITn};}\\
    	\scriptsize{$^c$\url{https://bit.ly/3nyVzvV}.}
    	\end{tablenotes}
	\end{threeparttable}
\end{table}

However, it is non-trivial to achieve real performance gain with top-k sparsification compared to the All-Reduce counterpart due to two main reasons: 1) the top-k selection is very inefficient on GPUs, and 2) the sparsified communication generally requires an All-Gather operation~\cite{renggli2019sparcml} which achieves very low performance on public GPU clusters. In addition, I/O with networked file systems (NFS) should also be carefully designed to achieve higher throughput on large-scale training due to the extensive data access at every training iterations. Table~\ref{table:public-cloud-instances} presents some popular public cloud instances which are with moderate storage and network connections on high-end computing servers. An experiment of training ResNet-50 with the ImageNet data set using TensorFlow and Horovod with their highly optimized configurations shows that 128 Nvidia V100 GPUs in Tencent Cloud (details in \S\ref{sec:experiments}) can only achieve about $40\times$ speedup compared to a single V100 GPU, which results in a very low scaling efficiency of $31\%$. We will study the training performance of existing state-of-the-art systems in \S\ref{sec:background}.

To this end, in this paper, we propose an efficient communication library with top-k sparsification, in which we first propose a novel approximate top-k operator that is friendly to the GPU architecture, and then we propose a hierarchical top-k communication to better utilize the connection bandwidth resources. To further improve the system scalability, we propose a simple yet effective mechanism for I/O reading with multi-level data caching and parallel gradient pro-processing for learning rates calculation. The technical contributions of this paper are summarized as follows: 
\begin{itemize}
    \item We propose an efficient approximate top-k gradient sparsification algorithm on GPUs to compress the communication data with very slight computation overheads.
    \item We present a novel hierarchical communication algorithm to aggregate the sparsified gradients to better utilize the bandwidth on the GPU clusters that are with fast intraconnects and slow interconnects.
    \item We design an efficient distributed training system for DNNs atop TensorFlow and Horovod, in which we propose a multi-level data caching mechanism and a generic parallel tensor operator to further improve the system scalability.
    \item We perform distributed training experiments with two types of models, CNNs and Transformer, to demonstrate the effectiveness of the proposed techniques on a Tencent Cloud cluster with 16 nodes connected with 25 Gbps Ethernet (each node has 8 Nvidia Tesla V100-32GB GPUs with NVLink). Experimental results show that our system achieves $25\%-40\%$ faster than existing state-of-the-art systems on CNNs and Transformer.
    \item We finally demonstrate a case study on the DAWNBench\footnote{\url{https://dawn.cs.stanford.edu/benchmark/ImageNet/train.html}} leader-board to achieve the fastest training time on ResNet-50 even with a slower inter-node connection than existing work.
\end{itemize}

The rest of the paper is organized as follows. We first introduce the background of distributed training of deep learning and demonstrate the limitations of existing solutions in \S\ref{sec:background}. Then we present our efficient communication library with top-k sparsification in \S\ref{sec:commlib}. We introduce our training system with technical details of our proposed novel approaches in \S\ref{sec:system}. The experimental studies are presented in \S\ref{sec:experiments}. We introduce the related work in \S\ref{sec:relatedwork}, and finally we conclude the paper in \S\ref{sec:conclusion}.

\section{Background and Motivation}\label{sec:background}
\subsection{Synchronous SGD with Data Parallelism}
Synchronous SGD with data parallelism iteratively updates the model parameters with Eq.~\eqref{equ:ssgd}. The training process of each iteration can be decoupled into several steps. 1) Each worker loads a mini-batch of training data from the file system, and the data is then pre-processed (e.g., decoding and argumentation) as the input for the deep model. 2) Each GPU performs feed-forward and backpropagation to computes the local gradients. 3) The local gradients are then aggregated through an All-Reduce operation such that all GPUs have consistent global gradients. 4) The global gradients are then used to compute the model updates along with the learning rate, e.g., layer-wise adaptive rate scaling (LARS)~\cite{you2018imagenet}.

Let $b$ and $B$ denote the local batch size and the global batch size of one training iteration respectively. For a $P$-worker cluster, $B=b\times P$. Assume that the throughput of the system is $T$, the number of epochs to train a model is $E$, and the total number of samples of the data set is $N$, we can represent the budget of training a model by $\frac{N\times E}{T}$. Therefore, given a $P$-worker cluster, we should try to reduce the iteration time $t_{iter}$ to increase the system throughput $T=\frac{b\times P}{t_{iter}}$, where $b$ is generally set to maximally occupy the GPU memory.

\subsection{Existing Problems on Public Clouds}\label{subsec:problems}
\begin{figure}[!ht]
	\centering
	\includegraphics[width=0.8\linewidth]{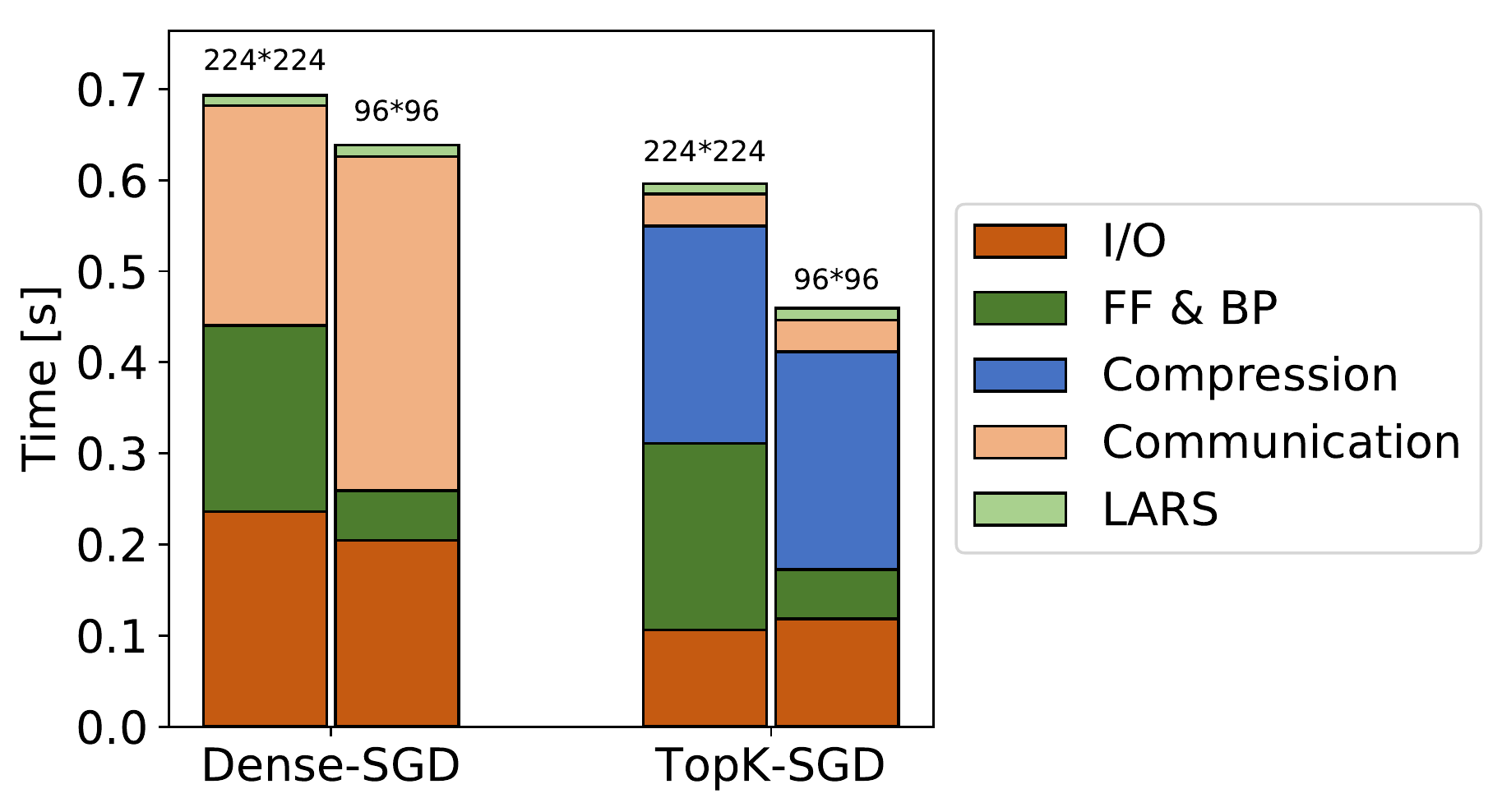}
	\caption{Time breakdown of one iteration with existing training schemes. FF\&BP indicates the feed-forward and backpropagation computations. The numbers on the top of the bars indicate the resolution of input images.}
	\label{fig:naiveresults}
\end{figure}

To demonstrate the training efficiency of existing state-of-the-art solutions on public clouds, we choose the highest configuration on Tencent Cloud with a 16-node GPU cluster connected with 25Gbps Ethernet (25GbE), where each node has 8 Nvidia V100 GPUs connected with NVLink, to measure the training performance of ResNet-50~\cite{he2016deep} on the ImageNet~\cite{deng2009imagenet} data set. As for large-batch training, layer-wise adaptive rate scaling (LARS)~\cite{you2018imagenet} or LAMB~\cite{you2020large} is required to preserve the model generalization ability, so we include the LARS computation in the training. We use Dense-SGD to denote the training scheme of synchronous SGD with the full dense gradients for aggregation. The results are shown in Fig.~\ref{fig:naiveresults}. Note that the feed-forward and backpropagation computing tasks, the gradient communication tasks, and the LARS computing tasks may be executed in parallel if possible (e.g., wait-free backpropagation~\cite{zhang2017poseidon,awan2017s} and tensor fusion~\cite{shi2019mg,shi2020communication}), so the time breakdown shown in Fig.~\ref{fig:naiveresults} is the elapsed time that cannot be overlapped by tasks pipelining.

From Fig.~\ref{fig:naiveresults}, we can observe that the I/O time and communication time occupy a large portion of the overall iteration time. For the gradient communication, gradient compression, such as top-k sparsification (TopK-SGD)~\cite{lin2018deep,renggli2019sparcml}, can reduce the communication traffic with little impact on the model accuracy. However, there exist two main problems in TopK-SGD, 1) it requires exact top-k selections on GPUs, which could be very slow with the naive implementation, and 2) the top-k elements in each GPUs might have different indices in the original dense gradients so that we need to use the All-Gather collective instead of All-Reduce to aggregate the gradients~\cite{renggli2019sparcml}. As shown in Fig.~\ref{fig:naiveresults}, TopK-SGD significantly reduces the communication time, but it introduces an extra top-k compression overhead, which is around 0.239 seconds, while the total of feed-forward and backpropagation time is only 0.204 seconds. On the cases with the small resolution input $96\times 96$, feed-forward and backpropagation time is very small so that the LARS computing time is also relatively significant compared with the feed-forward and backpropagation time. 

Therefore, it requires careful design to optimize gradient communication on public cloud clusters, and the I/O and LARS overheads can be also optimized to improve the overall system throughput.

\section{CommLib: An Efficient Gradient Communication Library}\label{sec:commlib}
As we demonstrated in Section~\ref{subsec:problems}, the communication cost is significant on the 16-node GPU cloud cluster. We propose a novel sparse gradient communication scheme, which contains two parts: 1) an approximate top-k operator that is friendly to the GPU architecture, and 2) a hierarchical top-k communication algorithm that can better utilize the bandwidth resources.

\subsection{MSTopK: an Approximate Top-k Operator}\label{subsec:mstopk}
The top-k sparsification requires a top-k selection on the gradient tensors which are stored on GPUs, which can be formally defined as follows~\cite{lin2018deep,alistarh2018convergence,shi2019convergence}. For any input vector $\bm{x}\in\mathbb{R}^d$, the top-k operator $\text{TopK}(\bm{x}, k)\in \mathbb{R}^d$ whose $i^{th}$ element is
\begin{equation}
    \text{TopK}(\bm{x}, k)^{(i)}=
    \begin{cases}
    \bm{x}^{(i)}, &\text{ if } |\bm{x}^{(i)}|>thres \\
    0, &\text{ otherwise}
    \end{cases},
\end{equation}
where $x^{(i)}$ is the $i^{th}$ element of $\bm{x}$, and $thres$ is the $k^{th}$ largest value of $|\bm{x}|$. Due to the irregular access of GPU memory in the top-k selection, it is non-trivial to implement an efficient top-k selection algorithm on GPUs~\cite{shanbhag2018efficient,lin2018deep}. To this end, we design an approximate top-k selection algorithm with multiple samplings, named MSTopK, which is friendly to many-core processors. The key idea of MSTopK is to use a binary search to find thresholds that are close to the exact threshold (say $thres$).

\begin{figure}[!ht]
	\centering
	\includegraphics[width=0.8\linewidth]{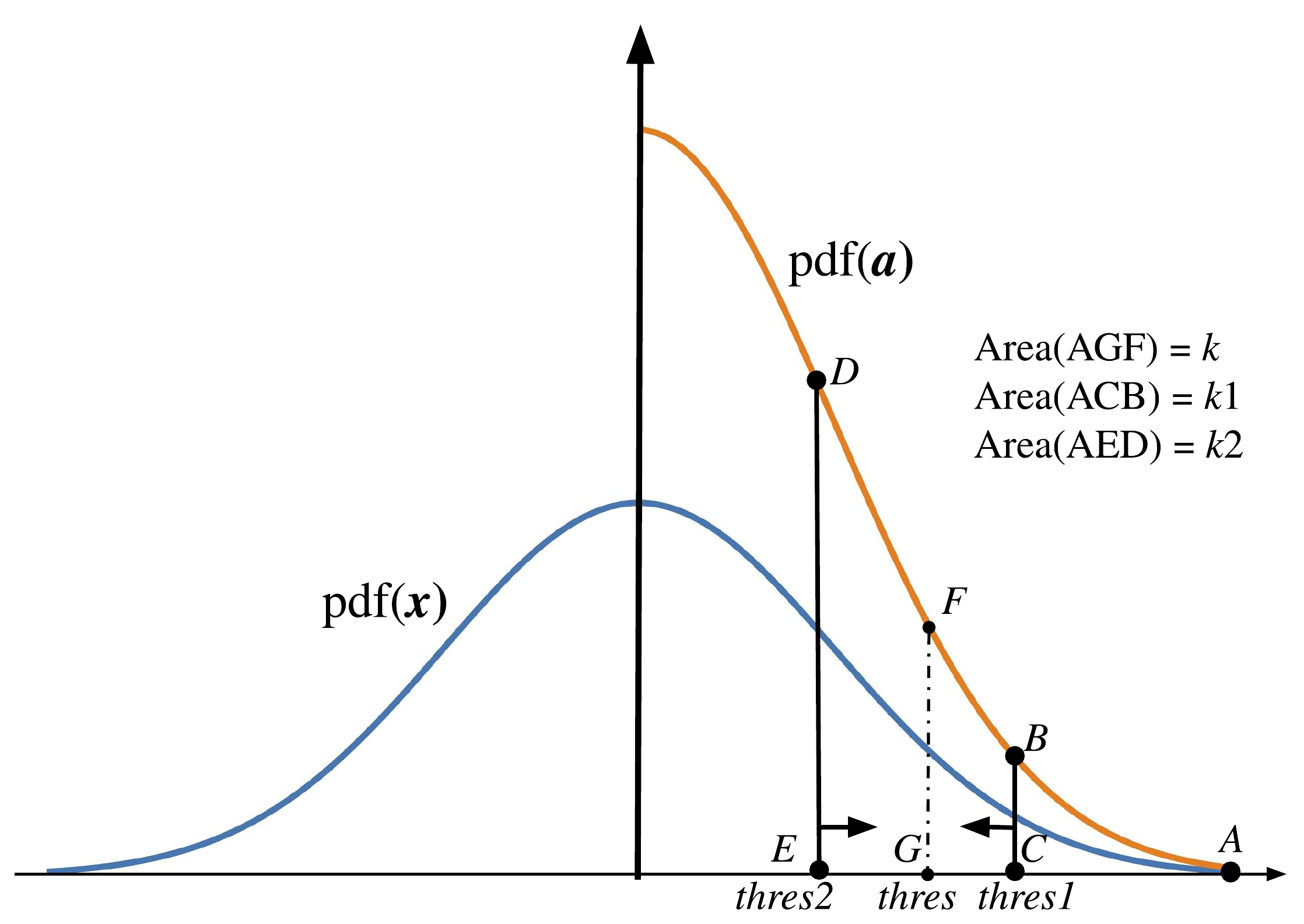}
	\caption{Illustration of the threshold search. Note that the exact threshold $thres$ is unknown, and we use the number of selected elements (e.g., $k1$ and $k2$) with each time's threshold to determine whether the chosen threshold is larger or smaller than the exact threshold. Every sampling of selection, we move $thres1$ and $thres2$ to be close to $thres$.}
	\label{fig:mstopk}
\end{figure}

Assume that we want to select $k$ elements from the input vector $\bm{x}\in \mathbb{R}^d$. We first use the average value ($\overline{a}$) of the absolute values of $\bm{x}$ data (say $\bm{a}=|\bm{x}|$) as the threshold ($thres1$) to select the elements (say $\bm{\kappa}$) whose values are not smaller than $\overline{a}$. If the dimension of $\bm{\kappa}$ is smaller (or larger) than $k$, then we half (or double) $thres1$ as $thres2$. If both $thres1$ and $thres2$ are smaller (or larger) than $thres$, we repeat the above search. After several trials, $thres$ should be located between $thres1$ and $thres2$ (as shown in Fig.~\ref{fig:mstopk}), then we can further narrow down the thresholds with the same pattern of search as previous. We set a fixed number of searches (say $N$), and finally we select $k$ elements using the chosen two thresholds. The pseudo-code of MSTopK is shown in Algorithm~\ref{algo:mstopk}. Note that there are no expensive memory access operations (e.g., sort) in Algorithm~\ref{algo:mstopk}, so it would be efficient on GPUs.

\begin{algorithm}[!ht]
\caption{MSTopK}\label{algo:mstopk}
\textbf{Input: }$\bm{x}\in\mathbb{R}^d, k, N$
\small
\begin{algorithmic}[1]
    \STATE $\bm{a}=\text{abs}(\bm{x})$;
    \STATE $\overline{a}=\text{mean}(\bm{a})$;
    \STATE $u=\max(\bm{a})$;
    \STATE $l=0;r=1;$
    \STATE $k1=0;k2=\text{len}(\bm{x})$;
    \STATE $thres1=0;thres2=0;$
    \FOR{$i=1\to N$}
        \STATE $ratio=l+(r-l)/2$;
        \STATE $thres=\overline{a}+ratio*(u-\overline{a})$;
        \STATE $nnz=\text{count\_nonzero}(\bm{a} \geq thres)$;
        \IF{$nnz\leq k$}
            \STATE $r=ratio$;
            \IF{$nnz>k1$}
                \STATE $k1=nnz$;
                \STATE $thres1=thres$;
            \ENDIF
        \ELSIF{$nnz>k$}
            \STATE $l=ratio$;
            \IF{$nnz<k2$}
                \STATE $k2=nnz;$
                \STATE $thres2=thres$;
            \ENDIF
        \ENDIF
    \ENDFOR
    \STATE $\bm{\iota}1=\text{nonzero\_indices}(\bm{a} \geq thres1)$;
    \STATE $\bm{\iota}2=\text{nonzero\_indices}((\bm{a}<thres1) \text{ and } (\bm{a} \geq thres2))$;
    \STATE $rand=\text{random}(0, \text{len}(\bm{\iota}2)-(k-k1)+1);$
    \STATE $\bm{\iota}=\text{concat}(\bm{\iota}1, \bm{\iota}2[rand:rand+k-k1]$;
    \STATE $\bm{\kappa}=\bm{x}[\bm{\iota}]$;
    \STATE Return $\bm{\kappa}, \bm{\iota}$;
\end{algorithmic}
\end{algorithm}

\subsection{Hierarchical Top-k Communication}\label{subsec:hitopcomm}
\begin{figure*}[!ht]
	\centering
	\includegraphics[width=\linewidth]{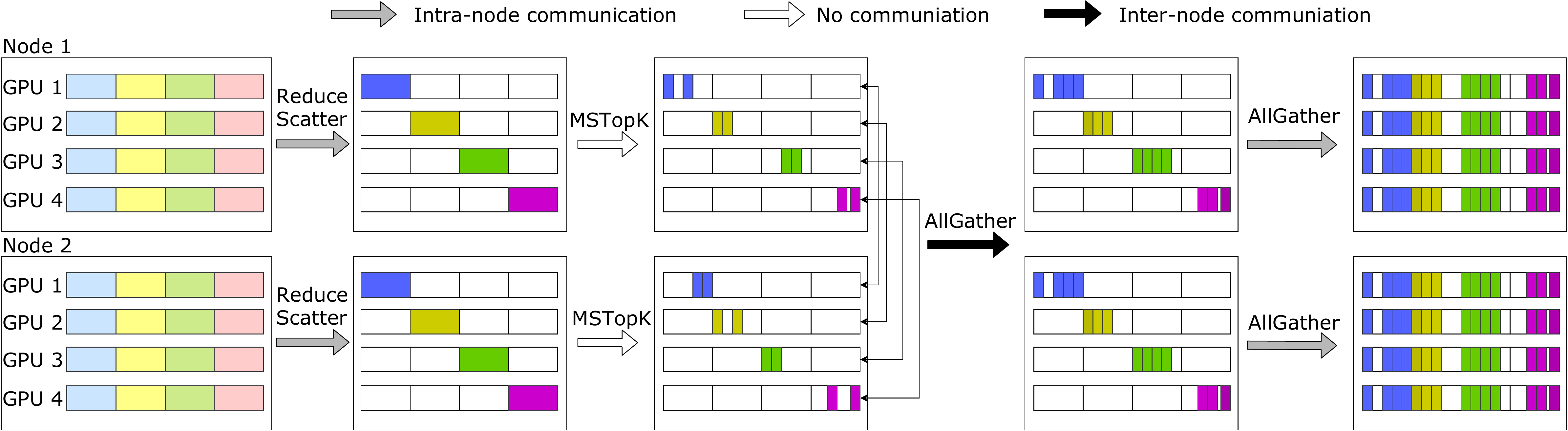}
	\caption{An example of hierarchical top-k communication with two nodes and each node has 4 GPUs.}
	\label{fig:hitopkcomm}
\end{figure*}
In top-k sparsification, the number of elements for worker $p$ to be transmitted becomes $2k$, which contains the vector of selected $k$ values, $\bm{v}^p$, and their corresponding indices, $\bm{i}^p$. $k=\rho \times d$, where $0< \rho < 1$ is called the density. Due to the irregular indices $\bm{i}^p$ at different workers, the values $\bm{v}^p$ cannot be aggregated through the All-Reduce collective. The efficient way is to use two All-Gather operations to aggregate the values and indices respectively~\cite{renggli2019sparcml}, and then the values are accumulated to the local gradients with the corresponding indices. However, each All-Gather operation for $P$ workers has a time complexity of 
\begin{equation}
    \alpha_{inter} \log P + 4(P-1)\beta_{inter} k,
\end{equation}
where $\alpha_{inter}$ is the latency of transmitting a message between two nodes, $\beta_{inter}$ is the transfer time per byte, and each element is represented by a 32-bit floating-point number (FP32). Note that $\beta_{inter}$ is the transmission speed between GPUs that are located in different nodes, which would be much slower than the intra-node transmission speed (e.g., NVLink). Therefore, directly using the All-Gather collective on a large-scale cluster connected with low-bandwidth and high-latency networks is very inefficient. To this end, we propose a hierarchical top-k communication (HiTopKComm) algorithm, which can better utilize both bandwidth resources of intra-node and inter-node connections.

Assume that the cluster has $m$ nodes and each node has $n$ GPUs, we use $\bm{g}_{i,j}\in \mathbb{R}^d$ to denote the local gradients at the $i^{th}$ node's $j^{th}$ GPU, where $1\leq i\leq m$ and $1\leq j \leq n$. Our HiTopKComm algorithm contains four steps as shown in Fig.~\ref{fig:hitopkcomm}. 1) All GPU nodes perform an intra-node Reduce-Scatter operation with $\bm{g}_{i,j}$ in parallel so that each GPU contains a $\frac{1}{n}$ summation of the $d$ elements, after which GPU $j$ at node $i$ has the updated gradients $\bm{g}_{i,j}^{[j]}\in\mathbb{R}^{d/n}$ and 
\begin{equation}
    \bm{g}_{i,j}^{[j]}=\bm{g}_{i,j}^{[(j-1)d/n, jd/n]}=\sum_{q=1}^n\bm{g}_{i,q}^{[(j-1)d/n, jd/n]}.
\end{equation}
2) Each GPU performs the top-k selection
\begin{equation}
    \bm{\kappa}_{i,j}, \bm{\iota}_{i,j}= \text{MSTopK}(\bm{g}_{i,j}^{[j]}, \rho\times d/n)
\end{equation}
using Algorithm~\ref{algo:mstopk}, which indicates that the MSTopK operation has a $n$ times smaller of the dimension of input data and the number of selected elements than selecting top-k elements from the original gradients $\bm{g}_{i,j}$. 3) Invoking $n$ communication streams for inter-node communications. That is, for the $j^{th}$ communication stream, the $j^{th}$ GPUs in all nodes perform an All-Gather operation with their $\bm{\kappa}_{i,j}$ and $\bm{\iota}_{i,j}$. As gradients from different GPUs in different nodes may have different indices, the gathered gradients should be accumulated with their corresponding indices, which results in a maximum of $\frac{\rho d}{n}m$ accumulated elements on each GPU. Formally
\begin{equation}
    \bm{g}_{i,j}^{[j]}=\sum_{p=1}^m \text{TopK}(\bm{g}_{i,j}^{[j]}, \rho d/n).
\end{equation}
4) All GPU nodes perform an intra-node All-Gather operation with $\bm{g}_{i,j}^{[j]}$ to construct $\bm{g}_{i,j}$. The pseudo-code of the HiTopKComm algorithm is shown in Algorithm~\ref{algo:hier-topk}.

\begin{algorithm}[!ht]
\caption{HiTopKComm}\label{algo:hier-topk}
\textbf{Input: }$\bm{g}_{i,j}\in\mathbb{R}^d, \rho, m, n$
\small
\begin{algorithmic}[1]
    \STATE Initiate $\tilde{\bm{g}}_{i,j}=[0]\in \mathbb{R}^d$; 
    \FOR{$i\in[m]$ in parallel}
        \STATE $\bm{g}_{i,j}$=Reduce-Scatter($\bm{g}_{i,j}$);
    \ENDFOR
    \STATE $\tilde{k}=\rho\times d/n$;
    \FOR{$i\in[m], j\in[n]$ in parallel} 
        \STATE $\bm{\kappa}_{i,j}, \bm{\iota}_{i,j}= \text{MSTopK}(\bm{g}_{i,j}^{[j]}, \tilde{k})$;
    \ENDFOR
    \STATE Initiate $\tilde{\bm{\kappa}}_{i,j}=[0]\in \mathbb{R}^{m\tilde{k}}$; 
    \STATE Initiate $\tilde{\bm{\iota}}_{i,j}=[0]\in \mathbb{N}^{m\tilde{k}}$; 
    \FOR{$j\in[n]$ in parallel} 
        \STATE $\tilde{\bm{\kappa}}_{i,j}$=All-Gather($\bm{\kappa}_{i,j}$);
        \STATE $\tilde{\bm{\iota}}_{i,j}$=All-Gather($\bm{\iota}_{i,j}$);
    \ENDFOR
    \FOR{$i\in[m], j\in[n]$ in parallel} 
        \FOR{$p=1\to m$}
            \STATE $\bm{\kappa}=\tilde{\bm{\kappa}}_{i,j}^{[p]}$; $\bm{\iota}=\tilde{\bm{\iota}}_{i,j}^{[p]}$;
            \STATE $\tilde{\bm{g}}_{i,j}[\bm{\iota}]+=\bm{\kappa}$;
        \ENDFOR
    \ENDFOR
    \FOR{$i\in[m]$ in parallel}
        \STATE $\tilde{\bm{g}}_{i,j}$=All-Gather($\tilde{\bm{g}}_{i,j}$);
    \ENDFOR
    \STATE Return $\tilde{\bm{g}}$;
\end{algorithmic}
\end{algorithm}

\textbf{Time complexity: } In HiTopKComm, the first step Reduce-Scatter is a ring-based algorithm, which takes a time complexity of 
\begin{equation}
    t_1^{HiTopKComm}=(n-1)\alpha_{intra}+\frac{4(n-1)d}{n}\beta_{intra},
\end{equation}
where $\alpha_{intra}$ and $\beta_{intra}$ are the latency and transfer time per byte two GPUs in a single node, respectively. Each element is represented by FP32. The second step is MSTopK with GPU computation, whose time complexity of is linear to the dimension of the input data and the selected number of elements, that is
\begin{equation}
    t_2^{HiTopKComm}\propto \mathcal{O}(\frac{d}{n}).
\end{equation}
The third step is the inter-node communication with an All-Gather operation, which has a time complexity of 
\begin{equation}
    t_3^{HiTopKComm}=\alpha_{inter} \log m + 4(m-1) \frac{\rho d}{n}\beta_{inter}.
\end{equation}
Finally, the last step is an intra-node All-Gather operation with the time complexity of
\begin{equation}
    t_4^{HiTopKComm}=\alpha_{intra} \log n + 4(n-1) \frac{\rho dm}{n}\beta_{intra},
\end{equation}
in which we assume the indices of the third step are all different so that the number of elements for All-Gather on each GPU in the last step is $\frac{\rho dm}{n}$. Since the inter-node communication is much slower than the intra-node communication, the main time-consuming part is the All-Gather operation in the third step.

\section{System Overview}\label{sec:system}
\begin{figure}[!ht]
	\centering
	\includegraphics[width=0.8\linewidth]{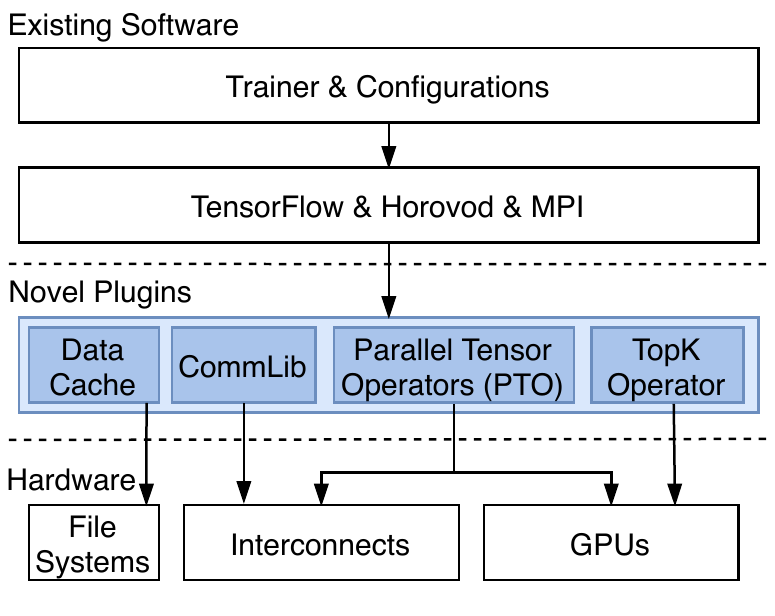}
	\caption{System overview with new proposed optimized plugins.}
	\label{fig:systemarch}
\end{figure}
We build our training system which integrates CommLib atop the widely used deep learning frameworks TensorFlow and Horovod as shown in Fig.~\ref{fig:systemarch}. Besides CommLib (with the approximate top-k operator), we propose two novel components to further improve the system scalability: 1) Data caching (DataCache), which provides automatically multi-level caching for efficient data reading, 2) Parallel tensor operator (PTO), which enables the computation on long tensors or a list of tensors be able to be computed on multiple GPUs in parallel.

\subsection{DataCache: Caching for Efficient Data Reading}\label{subsec:io}
\begin{figure}[!ht]
	\centering
	\includegraphics[width=0.8\linewidth]{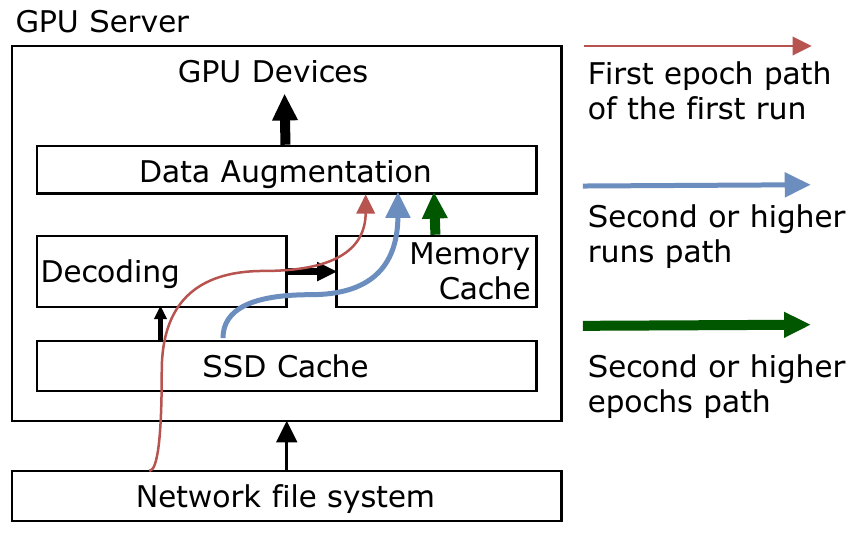}
	\caption{DataCache: Two-level caching for training data.}
	\label{fig:datacache}
\end{figure}
On the public clouds, training data is generally stored in a networked file system (NFS), whose reading performance may be limited by the network bandwidth and latency. On the other hand, training samples should be pre-processed before they are sent as input of the deep models. For example, in CNNs, the pre-processing process includes the decoding of input images (e.g., JPEG files) and normalization. Then the pre-processed data should be augmented (e.g., mirror, crop, etc.) before sent to GPU for training. To address the low performance of data access through the NFS and data pre-processing with CPU, we design a simple yet effective mechanism, DataCache, as shown in Fig.~\ref{fig:datacache}.

At the first epoch, the data should be read from the NFS into the local machine with the local file system cache. The data should be further decoded as the format that can be fed as the inputs of the deep models, which could consume many CPU resources. We further cache the pre-processed data into memory using the key-value store, where the key is the sample index and the value is the pre-processed data. In this way, the memory cache can significantly reduce the I/O time during the training process. To reduce memory consumption, the full data set is split into multiple parts that are separately stored on multiple nodes. Starting from the second epoch, all the data has been stored in the memory cache. With pipelining between data reading and GPU computations, the time cost of data reading from the memory cache can be almost fully overlapped by GPU computations. In summary, on one hand, enabling the local file system cache to store the data makes the multiple runs for hyper-parameter tuning be more efficient on data reading. On the other hand, using the memory cache makes the data reading from the second epochs at each run be fast.

\subsection{Parallel Processing of Tensors}\label{subsec:paralleltensors}
Processing tensors that are with the same data is very common in distributed training. In general, after the gradient aggregation, all GPUs have the same data of gradients and model parameters which should be further processed before update the model according to the optimizer. For example, the layer-wise adaptive rate scaling (LARS)~\cite{you2018imagenet} algorithm needs to calculate the layer-wise learning rates by 
\begin{equation}\label{equ:lars}
    \lambda_t^{(l)} = \gamma \times \eta_t \times \frac{\|\bm{w}_t^{(l)}\|}{\|\bm{g}_t^{(l)}\| + \epsilon\|\bm{w}_t^{(l)}\|},
\end{equation}
where $\gamma$ is a hyperparameter and $\epsilon$ is the weight decay. In the traditional training mechanism, all GPUs perform Eq.~\eqref{equ:lars} in parallel. Since the input and output are the same in all GPUs, we can partition the workload for different GPUs who first process different parts of data and then aggregate the generated results. We generalize this scheme as parallel processing tensors. 

For any tensor $\bm{g}\in \mathbb{R}^d$ that has been stored in all $P$ workers, if there is an operation $\text{OP}(\cdot)$ that regards $\bm{g}$ as the input and generates the same output, i.e.,
\begin{equation}\label{equ:sequential-op}
    r=\text{OP}(\bm{g}),
\end{equation}
then $\bm{g}$ can be partitioned into $P$ pieces and each GPU only computes one piece with OP. Formally, for $p=1,2,...,P$,
\begin{equation}\label{equ:parallel-op-1}
    r^{[p])}=\text{OP}(\bm{g}^{[p]})
\end{equation}
can be parallelized on $P$ GPUs, and the results are then aggregated by
\begin{equation}\label{equ:parallel-op-2}
    r=\text{All-Gather}(r^{[p]}).
\end{equation}
We use the parallel tensor operator (PTO) to denote the process of Eq.~\eqref{equ:parallel-op-1}~\eqref{equ:parallel-op-2}. Obviously, using PTO can reduce the computation workload on each GPU by $P$ times while it introduces an extra All-Gather communication overhead. In practice, if the time cost of the All-Gather operation is smaller than the time reduction of computing, PTO can accelerate the computation of Eq.~\eqref{equ:sequential-op}.

\textbf{PTO for LARS}: For the LARS computation as shown in Eq.~\eqref{equ:lars}, we should calculate the norms of each layer's gradients and weights to generate the layer-wise learning rates. We partition the workload in terms of the layer for different GPUs, which means different GPUs calculate different layers' learning rates that are finally gathered for every GPU. For example, in our 128-GPU experiments, the computations layer-wise learning rates for the ResNet-50 model, which has 161 layers, are distributed to the 128 GPUs. The first GPU calculates 1 to 2 layers' learning rates, the second one calculates layer 3 to 4, and so on. Finally, the layer-wise learning rates on the GPUs are all-gathered, which is with very low communication traffic as each layer's learning rate is a scalar. It would be similar to handle the case of LAMB~\cite{you2020large} using PTO.

\section{Experimental Studies}\label{sec:experiments}
In this section, we present the experimental studies to demonstrate the effectiveness of our proposed optimizations with real-world applications. We first show the detailed configuration of our testbed, and then demonstrate the experimental results on the efficiency of our CommLib, data caching, and PTO. After that, we compare the end-to-end training efficiency by putting the optimizations together. Finally, we present a case study in training ResNet-50 by breaking the record of DAWNBench\footnote{\url{https://dawn.cs.stanford.edu/benchmark/ImageNet/train.html}} in terms of the training time to the top-5 accuracy of 93\%.

\subsection{Experimental Environments}
\textbf{Testbed}: We choose the cluster from Tencent Cloud with 16 GPU instances, each instance is a virtual machine equipped with 8 Nvidia Tesla V100-32GB GPUs connected with NVLink. The 16 instances are connected with the virtual private connection on 25Gbps Ethernet (25GbE). The hardware and software are the same in all instances. The hardware is shown in Table~\ref{table:public-cloud-instances}, the OS in each instance is Linux-2.2, and the performance related libraries are: CUDA-10.1, cuDNN-7.6, NCCL-2.5.6, TensorFlow-1.15, and Hovorod-0.19.1.

\textbf{DNNs}: We choose two popular deep learning applications of computer vision using CNNs and natural language processing using Transformer. For CNNs, we choose ResNet-50 and VGG-19 on the ImageNet data set, while for Transformer, we choose Transformer\footnote{\url{https://bit.ly/34H7tLB}} from~\cite{vaswani2017attention} on the WMT17~\footnote{\url{https://bit.ly/34Epbzx}} data set.

\subsection{Top-k Operator Comparison}
\begin{figure}[!ht]
	\centering
		\begin{subfigure}{0.23\textwidth}
    	\includegraphics[width=\linewidth]{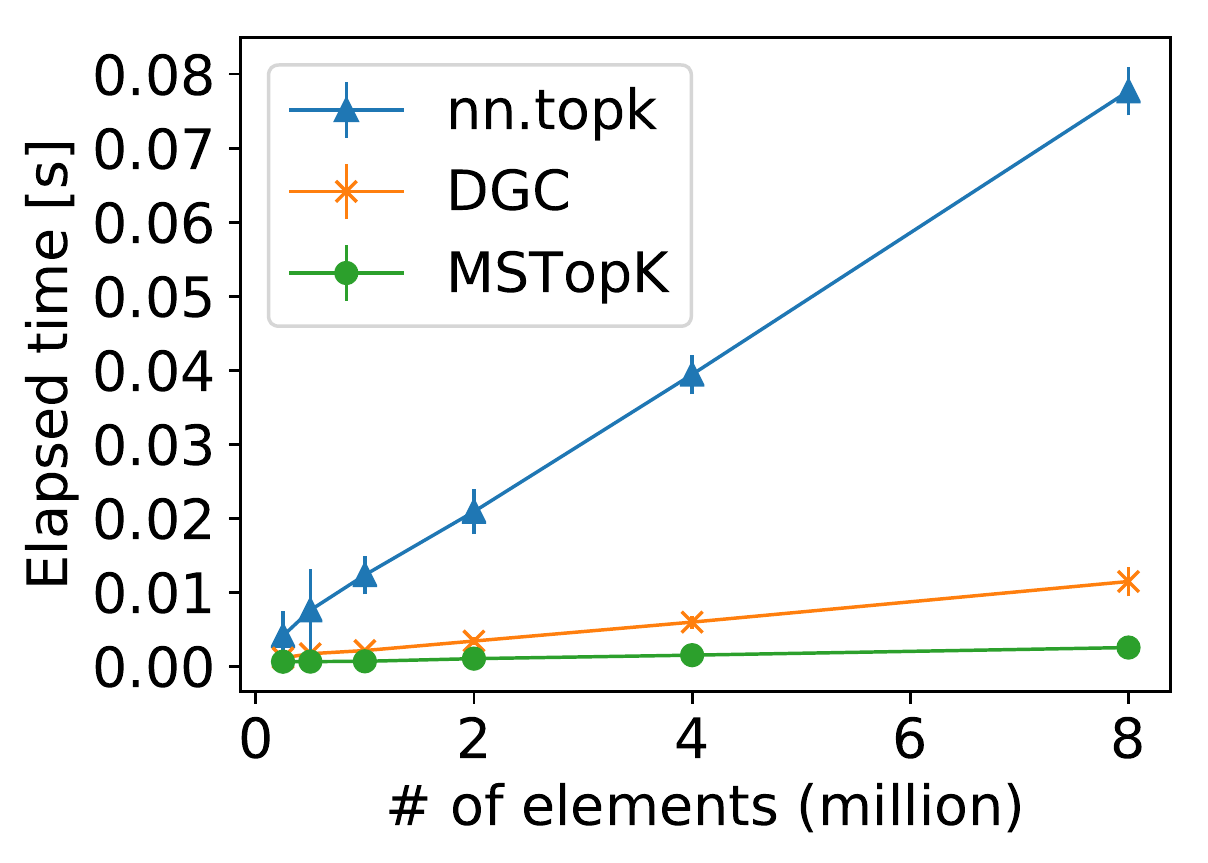}
    	\caption{Small tensors.}
	\end{subfigure}
	\begin{subfigure}{0.23\textwidth}
		\includegraphics[width=\linewidth]{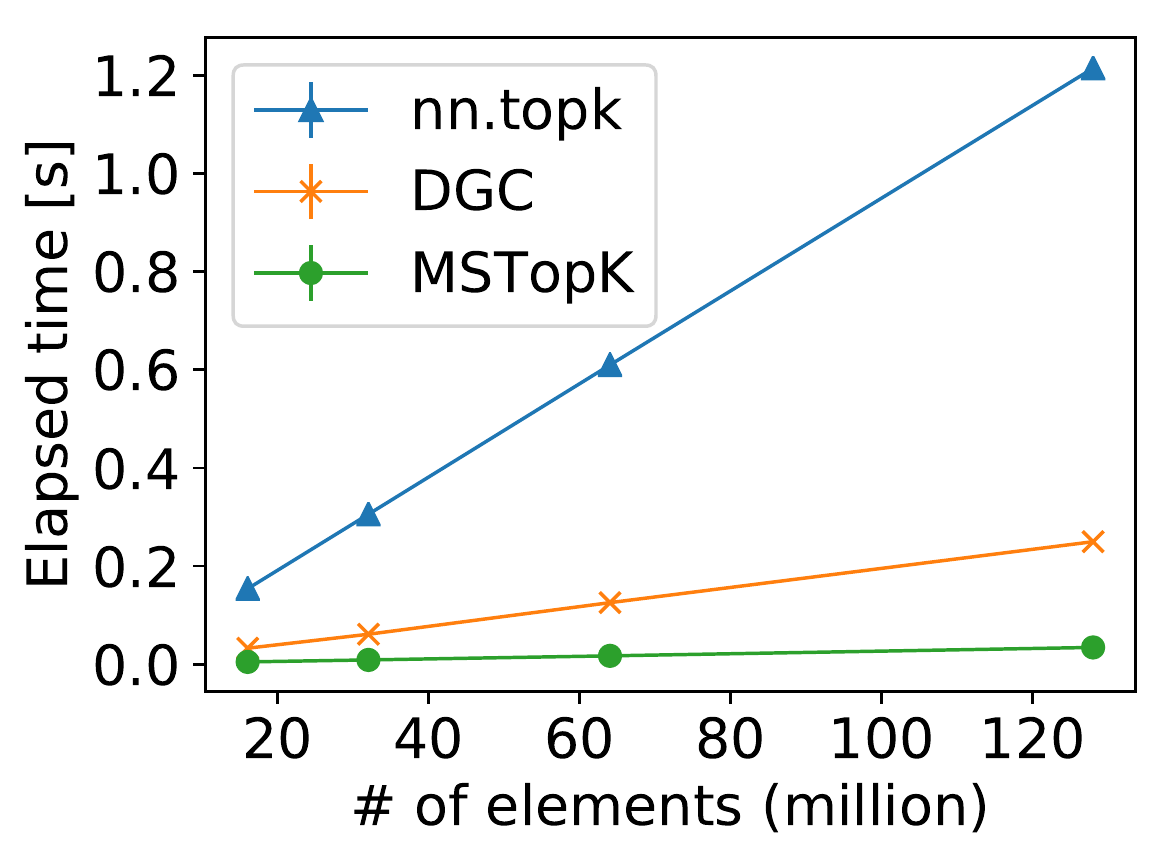}
		\caption{Large tensors.}
	\end{subfigure}
	\caption{Time performance between MSTopK, DGC~\cite{lin2018deep}, and nn.topk. The elapsed time is the average time of 5 independent experiments, and for each experiment, we run 5 warmup iterations and 100 iterations to measure the average. The number of samplings for MSTopK is 30.}
	\label{fig:mstopk-vs-nntopk}
\end{figure}
We compare the performance of our MSTopK operator with the naive top-k operator of TensorFlow (i.e., nn.topk) on a Tesla V100 GPU with different length of vectors from 256 thousand to 128 million. We also implement the top-k selection with double sampling in~\cite{lin2018deep}, which we denote as $DGC$. The number of selected elements is thousandths of the lengths of the vector, that is $k=0.001\times d$. The experimental results are shown in Fig.~\ref{fig:mstopk-vs-nntopk}. It can be seen that the exact top-k operator is very slow, while our MSTopK only requires a negligible computing time. The exact top-k selection on the GPU generally requires irregular memory access which is not friendly to the GPU architecture~\cite{shanbhag2018efficient,mei2016dissecting}. DGC is much better than the naive implementation, but it is still not fast enough as it also requires the exact top-k selections. Our MSTopK is an approximate operator that eliminates the irregular memory access using the multiple thresholds, which significantly improve the GPU memory access bandwidth with coalesced access of a large number of threads~\cite{cook2012cuda}. The results show that MSTopK significantly reduces the top-k selection time on GPUs.

\subsection{Performance of HiTopKComm}
\begin{figure}[!ht]
	\centering
	\centering
		\begin{subfigure}{0.23\textwidth}
    	\includegraphics[width=\linewidth]{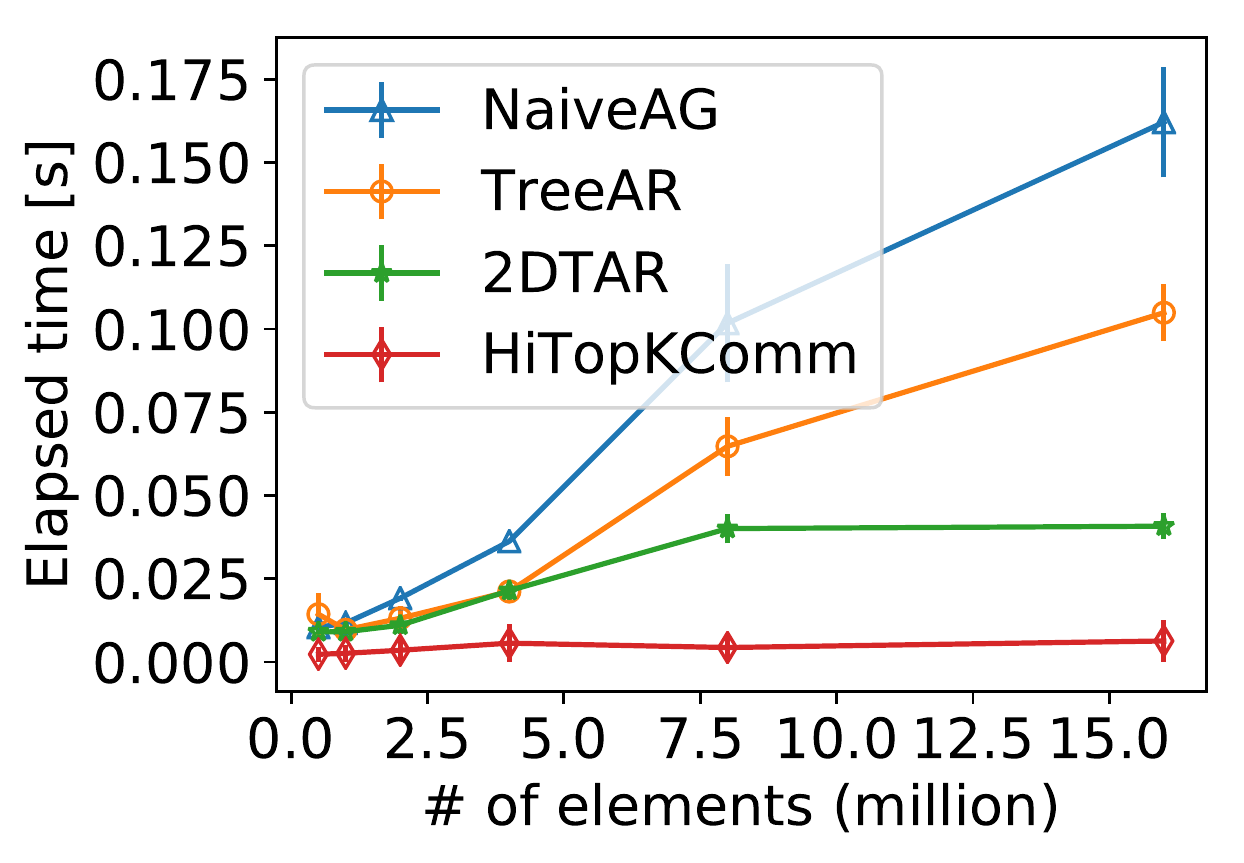}
    	\caption{Small tensors.}
	\end{subfigure}
	\begin{subfigure}{0.23\textwidth}
		\includegraphics[width=\linewidth]{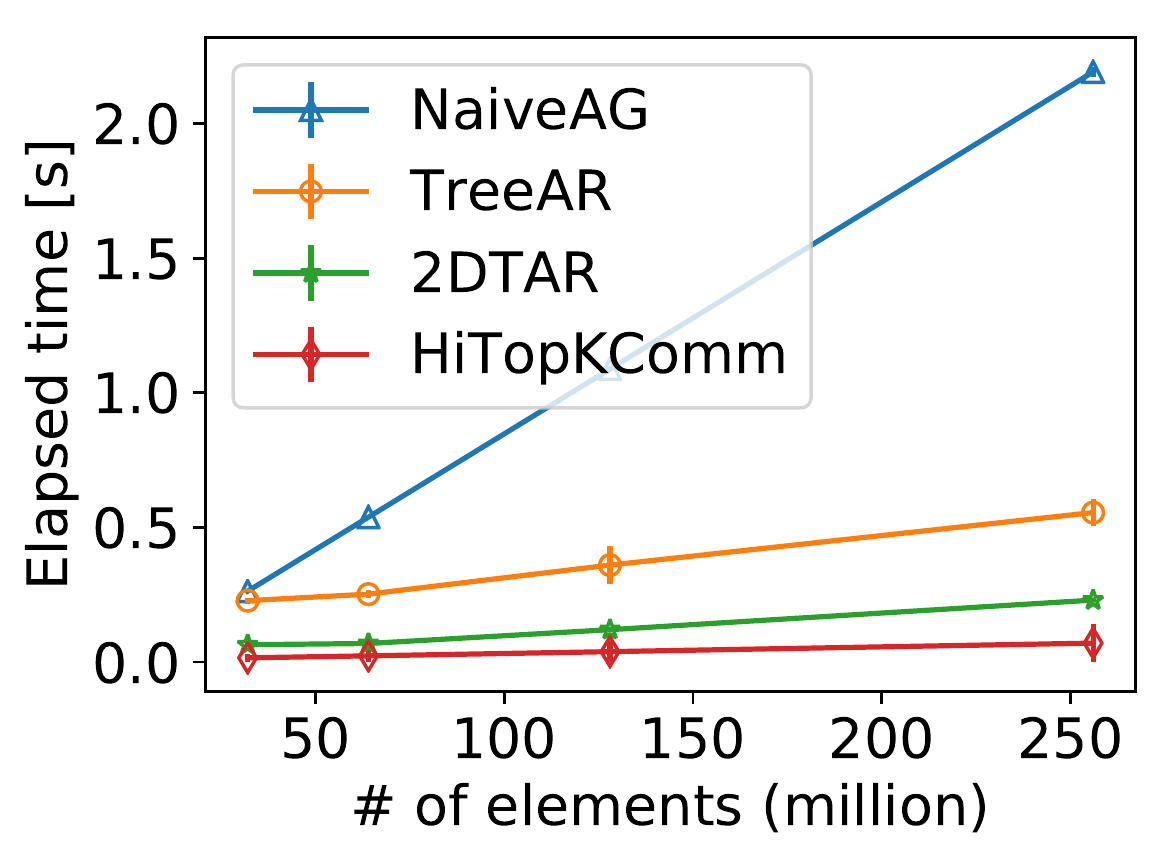}
		\caption{Large tensors.}
	\end{subfigure}
	\caption{Data aggregation time of different methods (NaiveAG, TreeAR, 2DTAR, and HiTopKComm). For the sparse communication, we use the density $\rho=0.01$.}
	\label{fig:hitopkcomm-perf}
\end{figure}

To show the effectiveness of our proposed HiTopKComm, we compare the communication efficiency with the original sparsified aggregation with All-Gather (NaiveAG) using NCCL and the tree-based All-Reduce (TreeAR) from NCCL. We also implement the 2D-Torus All-Reduce (2DTAR) algorithm~\cite{mikami2018massively,cho2019blueconnect} in our CommLib. 2DTAR can also exploit the hierarchical network connections to perform more efficient all-reducing. The results are shown in Fig.~\ref{fig:hitopkcomm-perf}. Note that we use the 16-bit floating point (FP16) for each element which is widely used in V100 GPU clusters. It is seen that NaiveAG is extremely inefficient due to the traditional All-Gather is not friendly to the cloud GPU clusters that are with imbalance bandwidth between intra-node and inter-node connections. For TreeAR which is highly optimized in NCCL, it is also not that efficient in the cloud environment. 2DTAR can better utilize the bandwidth resource to achieve efficient data communication. HiTopKComm, as expected, is the most efficient communication scheme among the four schemes we tested, which would help improve the system scalability.

\begin{figure}[!ht]
	\centering
	\begin{subfigure}{0.225\textwidth}
    	\includegraphics[width=\linewidth]{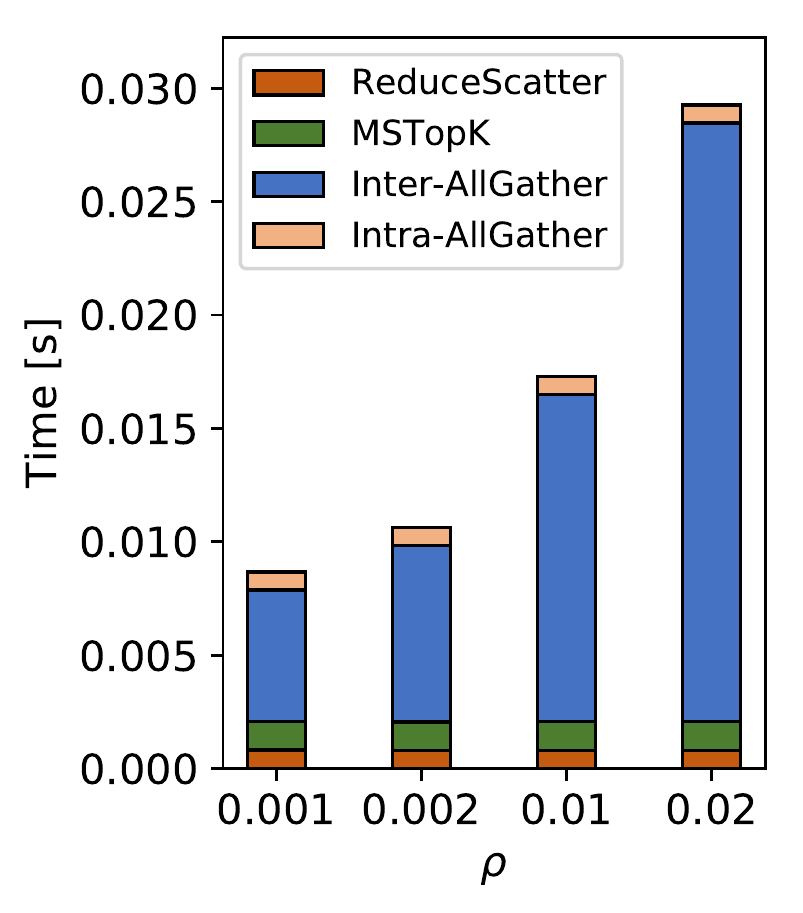}
    	\caption{ResNet-50.}
	\end{subfigure}
	\begin{subfigure}{0.22\textwidth}
		\includegraphics[width=\linewidth]{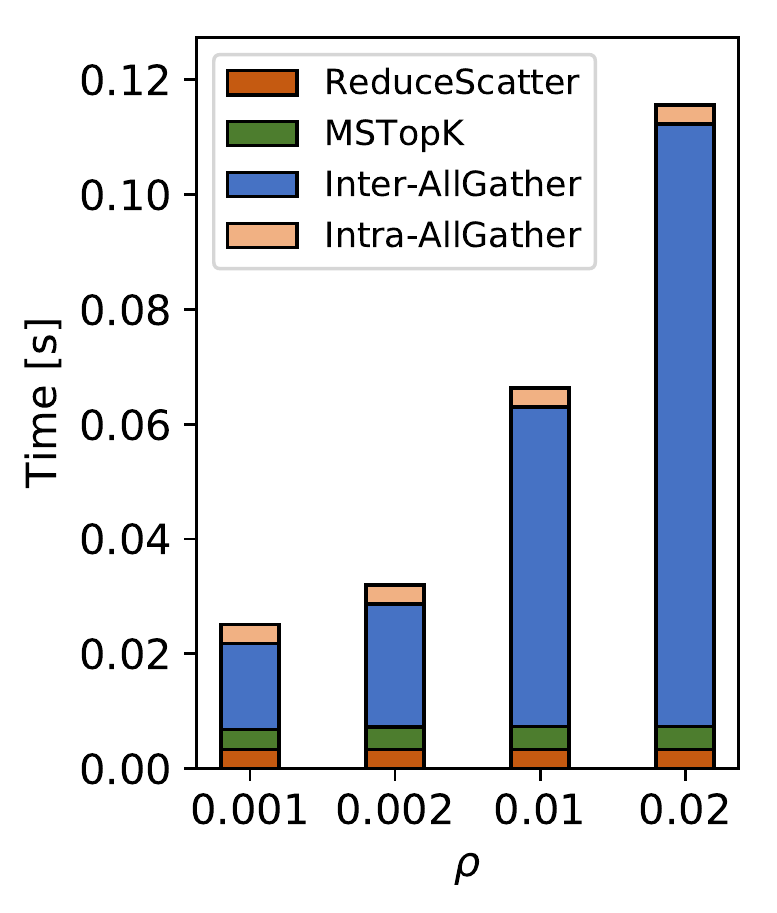}
		\caption{Transformer.}
	\end{subfigure}
	\caption{Time breakdown of HiTopKComm with different density.}
	\label{fig:hitopkcomm-breakdown}
\end{figure}

\textbf{Time Breakdown of HiTopKComm}. To understand the details of the efficiency of HiTopKComm, we breakdown the elapsed time for particular sizes of vectors according to the four steps (as shown in Fig.~\ref{fig:hitopkcomm}) of HiTopKComm. We use two specific cases (i.e., $25$ million parameters for ResNet-50 and $110$ million parameters for Transformer) using $k=0.01d$, both of which are with FP32 for each element. The results are shown in Fig.~\ref{fig:hitopkcomm-breakdown}. It is seen that the most time-consuming part is the inter-communication with the All-Gather operation due to the low bandwidth of inter-connection between multiple nodes. The time cost of top-k compression using MSTopK is very small, which is negligible. Due to the high-bandwidth and low latency of the intra-node connections between GPUs, the intra-node Reduce-Scatter and All-Gather operations also have a very small overhead.

\subsection{DataCache and PTO}
\begin{figure}[!h]
	\centering
	\includegraphics[width=0.45\linewidth]{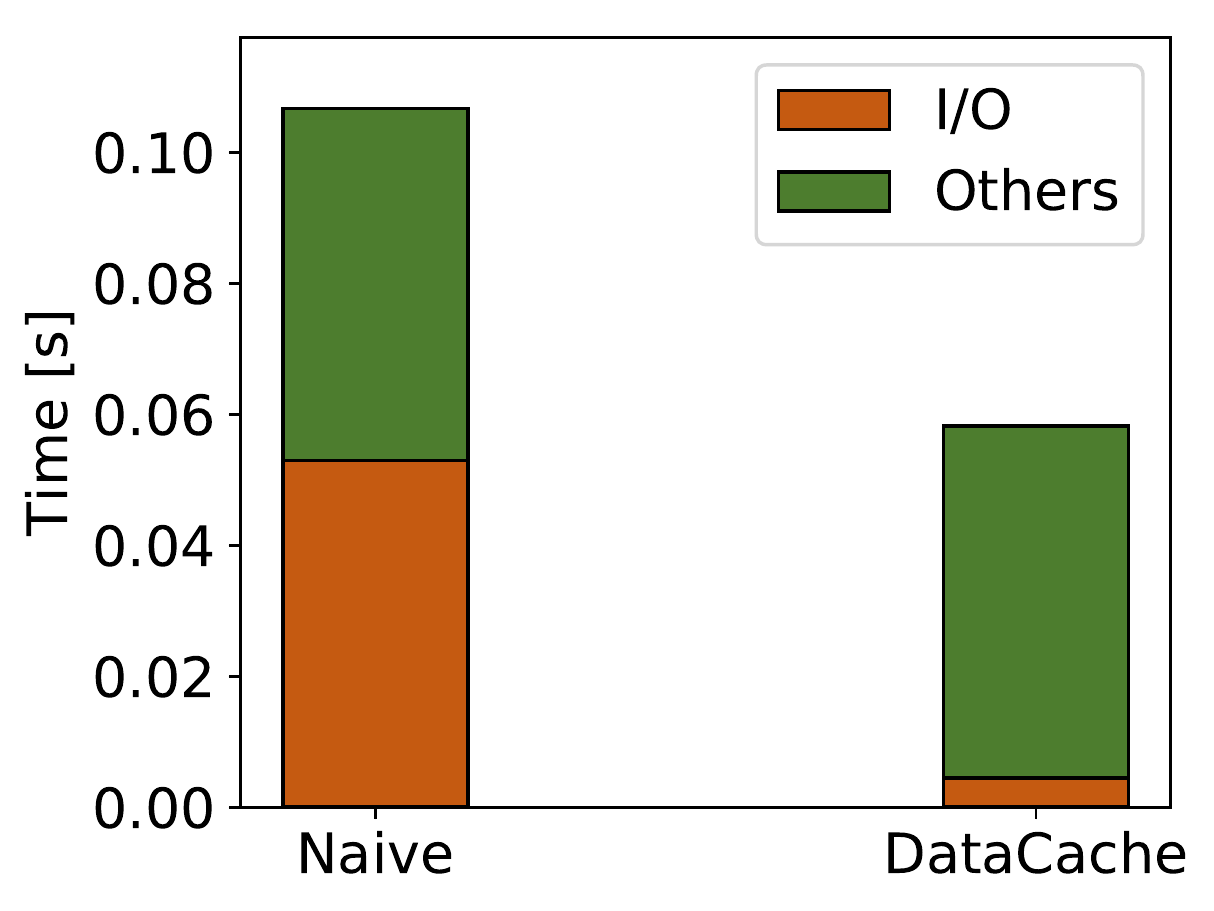}
	\caption{The performance comparison of training w/o (Naive) and w/ (DataCache) data caching. Using a V100 GPU on ResNet-50 with the input resolution of $96\times 96$.}
	\label{fig:io-perf}
\end{figure}
To show the benefits of data caching during the training process, we measure the end-to-end iteration time w/o and w/ our data caching. The result is shown in Fig.~\ref{fig:io-perf}. The I/O time is reduced over 10 times by using our DataCache, and the end-to-end training throughput is improved about 2 times.

To compare the performance of PTO, we also choose ResNet-50 and Transformer models to calculate LARS of Eq.~\eqref{equ:lars} with randomly generated $\bm{w}^{(l)}$ and $\bm{g}^{(l)}$. The original LARS computing time on ResNet-50 and Transformer is 11ms and 30ms respectively, while our PTO takes 7ms and 14ms respectively. Our PTO achieves about $2\times$ speedups on our 128 GPU cluster on both ResNet-50 and Transformer. 

\subsection{End-to-end Training}
Putting the overall optimizations together, we would like to evaluate the end-to-end training time improvement over the existing solutions. Since our MSTopK is an approximate top-k selection operator which may generate different results compared to the exact top-k operator, we first compare the convergence of our MSTopK solution on the large-scale training, and then we compare the end-to-end training efficiency on the 128 GPU cluster.

\subsubsection{Convergence}\label{subsubsec:convergence}
\begin{figure}[!ht]
	\centering
	\begin{subfigure}{0.23\textwidth}
    	\includegraphics[width=\linewidth]{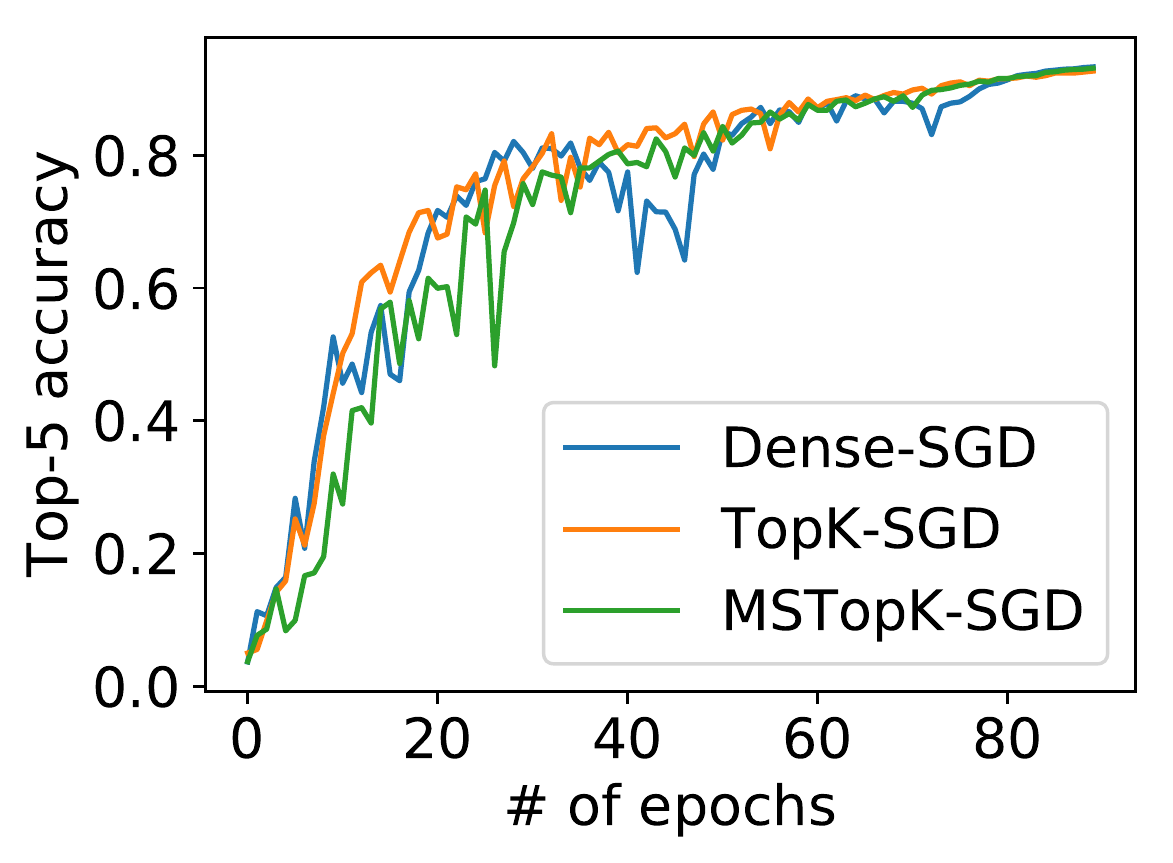}
    	\caption{ResNet-50.}
	\end{subfigure}
	\begin{subfigure}{0.23\textwidth}
		\includegraphics[width=\linewidth]{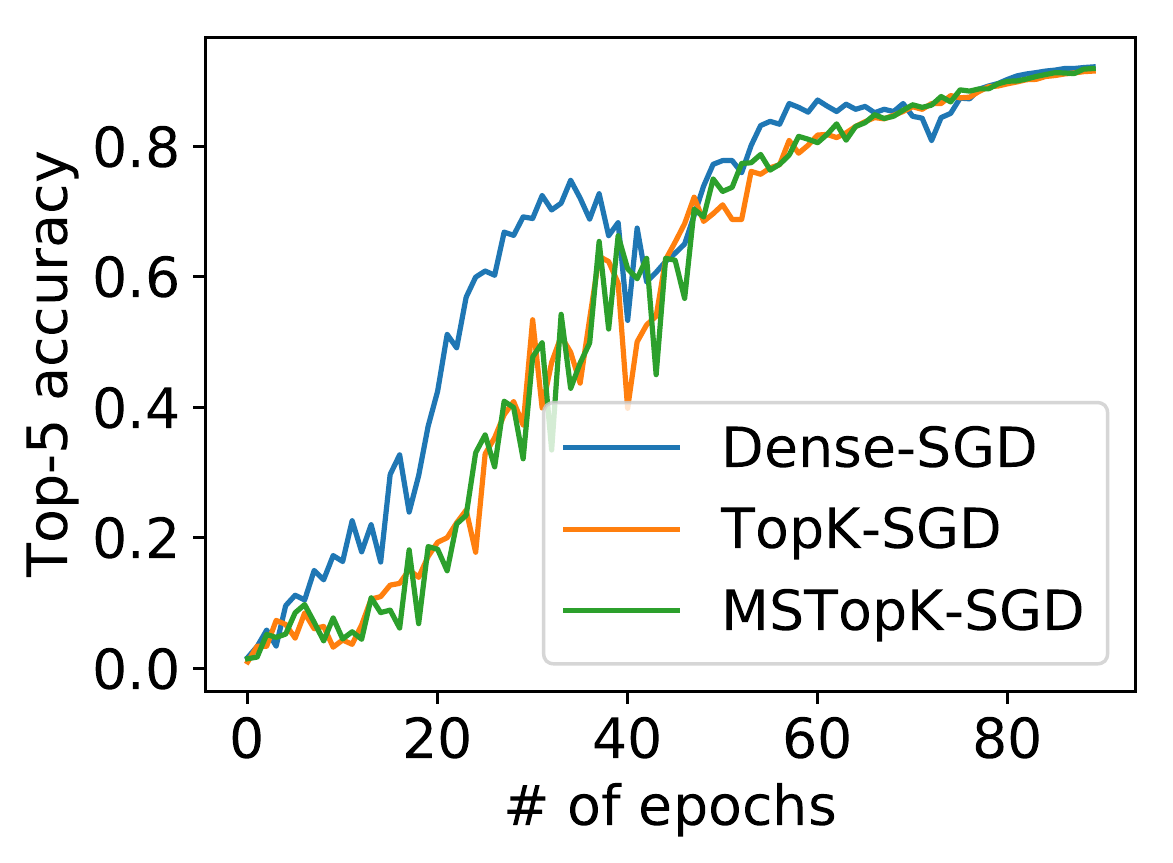}
		\caption{VGG19.}
	\end{subfigure}
	\caption{Convergence comparison.}
	\label{fig:convergence}
\end{figure}
For the convergence evaluation, we measure two kinds of applications (CNNs and Transformer) using three training algorithms of the original dense version (Dense-SGD) with TreeAR, the original top-k sparsification (TopK-SGD), and our proposed top-k compression (MSTopK-SGD). For CNNs, we train the ResNet-50 and VGG-19 models on the 128 GPU cluster in 90 epochs with the standard input resolution of $224\times 224$ and the local batch size is 256 (i.e., the global batch size is 32K). The validation accuracy during the training process is shown in Fig.~\ref{fig:convergence}. 

\begin{table}[!ht]
	\centering
	\caption{Validation performance (top-5 accuracy for CNNs and BLEU for Transformer).}
	\addtolength{\tabcolsep}{-2.5pt}
	\label{table:validation}
	\begin{tabular}{|c|c|c|c|}
	\hline
	 Model & 2DTAR-SGD & TopK-SGD & MSTopK-SGD \\\hline\hline
	 ResNet-50 & $93.31\%$ & $92.68\%$ & $93.12\%$ \\
	 VGG-19 & $92.19\%$ & $91.55\%$ & $91.94\%$ \\\hline
	 Transformer & $26.74$ & $24.42$ & $24.16$\\\hline
	\end{tabular}
\end{table}

In terms of evaluation performance, for CNNs, we use top-5 validation accuracy on the validation data set, and for Transformer~\cite{vaswani2017attention}, we measure BLEU on the validation data set by training 22k steps. The validation performance is shown in Table~\ref{table:validation}. Both the naive top-k sparsification and our MSTopK-SGD have slight accuracy loss compared to the dense version, which is reasonable due to the slower convergence speed of top-k sparsification than the dense version if the number of iterations is not large enough~\cite{karimireddy2019error}. 
Due to dense aggregation of gradients within intra-node, our MSTopK-SGD achieves slightly better than TopK-SGD on CNNs.

\subsubsection{Training Efficiency}
\begin{table*}[!ht]
	\centering
	\caption{System throughput and scaling efficiency of different SGD algorithms.}
	\addtolength{\tabcolsep}{-1pt}
	\label{table:scaling}
	\begin{tabular}{|c|c|c|c||c|c|c|}
	\hline
	 \multirow{2}{*}{Model} & \multicolumn{3}{c||}{System Throughput (samples/s)} & \multicolumn{3}{c|}{Scaling Efficiency (\%)}  \\\cline{2-7}
	       & Dense-SGD & 2DTAR-SGD & MSTopK-SGD & Dense-SGD & 2DTAR-SGD & MSTopK-SGD \\\hline\hline
	 ResNet-50 (224*224) & 64000 & \textbf{134656} & 133376 & 43.5 & \textbf{91.4} & 90.6 \\
	 ResNet-50 (96*96) & 113280 & 313600 & \textbf{396800} & 20.1 & 56.7 & \textbf{70.5} \\
	 VGG-19 & 17920 & 47616 & \textbf{57600} & 25 & 66.4 & \textbf{80.4} \\\hline
	 Transformer & 678 & 2534 & \textbf{3502} & 16.5 & 61.6 & \textbf{87.8} \\\hline
	\end{tabular}
\end{table*}
We compare the end-to-end training efficiency on the 128 GPU cluster and their scaling efficiency over the one GPU performance. Note that we enable the mixed-precision training technique so that the tensor cores of V100 GPUs can be used. The baseline throughput of single-GPU on ResNet-50, VGG-19, and Transformer are 1150, 560, and 32 samples/s, respectively. One sample indicates one image in CNNs, while it indicates one sentence with 256 words in Transformer. On our 128 GPU cluster, the performance comparison is shown in Table~\ref{table:scaling}. The results show that Dense-SGD with TreeAR is very inefficient. For 2DTAR-SGD, it is slightly faster than our MSTopK-SGD in the case of ResNet-50 with the input resolution of 224*224 because the computing time is long enough to overlap some communication costs. In all other cases, our MSTopK-SGD achieves 25\%-40\% improvement over 2DTAR-SGD.

\subsection{A Case Study: Break the Record of DAWNBench}
To further verify the effectiveness of our training system with the proposed novel optimizations, we present a case study on breaking the training time record of the ImageNet dataset on the leader-board of DAWNBench, which requires to achieve 93\% top-5 accuracy on $100,000$ validation samples. Existing top-4 leaderships on DAWNBench all exploit 128 Tesla V100 GPUs with 100Gbps InfiniBand (100GbIB) interconnect or 32Gbps Ethernet (32GbE), and the current fastest training speed on DAWNBench is 158 seconds by the team in Aliyun who exploited a 16-node GPU cluster (also with 128 V100 GPUs) connected with 32GbE. However, we exploit the Tencent Cloud cluster that is also with 16 GPU nodes while the interconnect between nodes is 25GbE, which is more challenging than running on Aliyun.

In recent practice, 93\% top-5 accuracy is achievable within only about 30 epochs using dynamic input image size during training. For example, in the current top-1 leadership on DAWNBench, the Alibaba team trained the model to achieve 93\% top-5 accuracy on the ImageNet data set with 28 epochs, where the first 13 epochs are with $96\times 96$ input resolution, the second 11 epochs with $128\times 128$ input resolution, the third 3 epochs with $224\times 224$ input resolution, and the last epoch with $288\times 288$ input resolution. On one hand, the small input size can achieve higher system throughput, but it makes the scaling be more difficult. On the other hand, the large input size has smaller system throughput, but it is helpful to achieve high validation accuracy. In the ImageNet training, it is known that the warmup process is necessary to preserve the model accuracy~\cite{goyal2017accurate,you2018imagenet,jia2018highly}, so recent studies~\footnote{\url{https://bit.ly/33KVbCO}} notice that in the first 10-20 epochs smaller input size can also warmup the training while achieving higher system throughput. We follow the 28-epoch training practice to 93\% top-5 accuracy on our 16-node cluster (128 V100 GPUs) with lower bandwidth interconnect (i.e., 25GbE).

\begin{table}[!ht]
	\centering
	\caption{System throughput (samples/s) with different input resolutions. BS indicates the local batch size at each GPU, and SE is shorten for scaling efficiency of the 128-GPU system.}
	\label{table:throughput}
	\addtolength{\tabcolsep}{-2.5pt}
	\begin{tabular}{|c|c|c|c|c|}
	\hline
	\# Epochs & Input & BS & Single-GPU & 128-GPU (SE)  \\\hline
	13 & $96\times 96$ & 256 & 4400 & 366,208 (65\%)  \\\hline
	11 & $128\times 128$ & 256 & 3010 & 269,696 (70\%) \\\hline
	3 &  $224\times 224$ & 256 & 1240 & 131,712 (83\%) \\\hline
	1 & $288\times 288$ & 128 & 710 & 72,960 (80\%) \\\hline
	\end{tabular}
	\vspace{-4pt}
\end{table}

In the first 13 epochs, the computing time of $96\times 96$ input is short on the V100 GPU, which makes the scaling efficiency very low using dense gradient aggregation. Therefore, we use MSTopK-SGD to train the model in the first 13 epochs, which exploits HiTopKComm for gradient aggregation to achieve higher system scaling efficiency. After that, we switch to use 2DTAR-SGD to balance the convergence speed and the system throughput. We cannot fully use MSTopK-SGD in the whole of 28 epochs because it would cause accuracy loss. Since when the input resolution is not smaller $128\times 128$, the scaling efficiency is acceptable, we start to use dense gradient communication so that we can achieve 93\% top-5 accuracy in 28 epochs. The training throughput with different input resolution is shown in Table~\ref{table:throughput}, and the training time using 128 Nvidia V100 GPUs to achieve 93\% validation accuracy of the ImageNet data set is shown in Table~\ref{table:28epochs}. The results show that our method achieves faster training time even with slower interconnects between GPU nodes.

\begin{table}[!ht]
\vspace{-4pt}
	\centering
	\caption{Time to 93\% top-5 accuracy with 128 Tesla V100 GPUs.}
	\label{table:28epochs}
	\begin{tabular}{|c|c|c|c|}
	\hline
	Team & Date & Interconnect & Time (seconds) \\\hline
	FastAI & Sep 2018 & 100GbIB & 1086  \\\hline
	Huawei & Dec 2018 & - & 562 \\\hline
	Huawei & May 2019 & 100GbIB & 163 \\\hline
	Alibaba & Mar 2020 & 32GbE & 158 \\\hline
	\textbf{Ours} & Aug 2020 & \textbf{25GbE} & \textbf{151} \\\hline
	\end{tabular}
\end{table}

\section{Related Work}\label{sec:relatedwork}
There exists much work trying to increase the scaling efficiency of large-scale distributed training while preserving the convergence properties.

In terms of gradient aggregation, there exist many algorithms to increase communication efficiency and gradient compression algorithms to reduce the communication traffic. For example, double binary trees~\cite{sanders2009two} All-Reduce has been integrated in NCCL to support large-scale communication in the HPC environment. Some All-Reduce algorithms~\cite{goyal2017accurate,jia2018highly,mikami2018massively,cho2019blueconnect,wang2020blink,luo2020plink,dong2020eflops,chu2020nv} have been proposed to efficiently all-reducing data in different environments. To reduce the communication traffic, gradient compression techniques including quantization and sparsification are widely studied~\cite{alistarh2017qsgd,lin2018deep,shi2019convergence,karimireddy2019error,shi2019distributed,vogels2019powersgd}.

Large-batch training is an effective way to improve the system scalability by increasing the workload of GPUs and thus reducing the communication-to-computation ratio, but it requires careful hyperparameter tuning to preserve the model performance. LARS~\cite{you2018imagenet} and LAMB~\cite{you2020large} are two main techniques to address the convergence problem in large-batch training. AdaScale SGD~\cite{johnson2020adascale} is another technique to make large-batch training be easier to converge.

For the GPU computing, most tensor operations highly rely on CUDA libraries like cuDNN, cuBLAS, etc. Most operators in training deep models are well supported by cuDNN, but the top-k operator is extremely inefficient. \cite{shanbhag2018efficient} tried to optimize the top-k query in the database area. The double sampling technique was proposed in~\cite{lin2018deep} to reduce the number of elements for the top-k operator, but it also requires at least two times of top-k operations on GPUs. In~\cite{fang2019redsync,shi2019understanding}, the authors proposed multiple sampling methods to approximate the top-k operation, but they may result in a different number of elements at different workers, which could make the data aggregation be inefficient.

Regarding I/O optimizations, \cite{pumma2019scalable} identified the inefficiency of data access for multi-GPU training, but they mainly focused on the problem of LMDB data that is relatively old in the Caffe framework. Recently, \cite{zhang2020efficient} proposed FanStore for efficient I/O that is particularly used on supercomputers.

\section{Conclusion}\label{sec:conclusion}
In this paper, we proposed a novel gradient communication library for distributed training systems, and we also optimize I/O and GPU computation to further improve the system scalability. Specifically, first, for I/O, we proposed a multi-level data caching scheme to reduce the I/O time on public clouds that generally use low-bandwidth NFS for data storage. Second, to improve the GPU efficiency on some operators, we design a GPU-friendly top-k sparsification operator and a parallel tensor operator that can better utilize multi-GPU computing power. Third, we proposed a hierarchical top-k communication for sparsified gradients. Extensive experiments were conducted on a 128 Tesla V100 GPU cluster (16 nodes) with 25Gbps Ethernet. Experimental results showed that our system outperforms the existing state-of-the-art solutions and breaks the record on training ResNet-50 to 93\% top-5 accuracy on DAWNBench.



\bibliography{main.bbl}
\bibliographystyle{mlsys2020}

\appendix
%


\end{document}